\documentclass{IEEEtran}
\usepackage{cite}
\usepackage{amsmath,amssymb,amsfonts}
\usepackage{float}
\usepackage{graphicx}
\graphicspath{{./figures/}}
\usepackage[caption=false,font=normalsize,labelfont=sf,textfont=sf]{subfig}
\usepackage{textcomp}
\usepackage{booktabs}
\usepackage{tikz}
\def\BibTeX{{\rm B\kern-.05em{\sc i\kern-.025em b}\kern-.08em
    T\kern-.1667em\lower.7ex\hbox{E}\kern-.125emX}}

\begin{document}
\title{Mitigation of Resonances in Sinuous Antennas}
\author{Dylan A. Crocker \IEEEmembership{Student, IEEE} and Waymond R. Scott, Jr., \IEEEmembership{Fellow, IEEE}
\thanks{This work was supported by the U.S. Army REDCOM CERDEC Night Vi- sion and Electronic Sensors Directorate, Science and Technology Division, Countermine Branch; the U.S. Army Research Office under Grant Number W911NF-11-1-0153; and by Sandia National Laboratories, a multimission laboratory managed and operated by National Technology and Engineering Solutions of Sandia, LLC., a wholly owned subsidiary of Honeywell International, Inc., for the U.S. Department of Energys National Nuclear Security Administration under contract DE-NA0003525. This paper describes objective technical results and analysis. Any subjective views or opinions that might be expressed in the paper do not necessarily represent the views of the U.S. Department of Energy or the United States Government.}
\thanks{D. A. Crocker is with Sandia National Laboratories, Albuquerque, NM 87123 USA (e-mail: dylan.crocker@sandia.gov). }
\thanks{W. R. Scott, Jr., is with the Georgia Institute of Technology, Atlanta, GA 30332 USA (e-mail:  waymond.scott@ece.gatech.edu).}
}

\maketitle

\begin{tikzpicture}[remember picture,overlay]
\node[anchor=137,rectangle,draw=white, inner sep=20pt] at (current page.120) {\Large SAND-2020-3198 J};
\end{tikzpicture}

\begin{abstract}
Sinuous antennas are capable of producing ultra-wideband radiation with polarization diversity. Such a capability makes the sinuous antenna an attractive candidate for wideband polarimetric radar applications. Additionally, the ability of the sinuous antenna to be implemented as a planar structure makes it a good fit for close in sensing applications such as ground penetrating radar. However, recent literature has shown the sinuous antenna to suffer from resonances which degrade performance. Such resonances produce late time ringing which is particularly troubling for pulsed close in sensing applications. The resonances occur in two forms: log-periodic resonances on the arms, and a resonance due to the outer truncation of the sinuous antenna geometry. A detailed investigation as to the correlation between the log-periodic resonances and the sinuous antenna design parameters indicates the resonances may be mitigated by selecting appropriate parameters. In addition, the resonance due to truncation may be mitigated by moving the circular truncation to the tip of the outermost cell which has advantages over the clipping method proposed in the literature.   
\end{abstract}


\section{Introduction}
\label{sec:introduction}
\IEEEPARstart{T}{he} sinuous antenna was first published in a patent by DuHamel in 1987. The patent describes the sinuous antenna as the combination of frequency independent spiral and log-periodic antenna concepts which resulted in a design capable of producing ultra-wideband (UWB) radiation with polarization diversity \cite{duhamel1987dual}. Such attributes have made the sinuous antenna useful in direction finding \cite{zhang2014four, bellion2008a}, human health monitoring \cite{xu2016design}, radio astronomy \cite{gawande2009gt, gawande2011towards, ivashina2014system, devilliers2017initial}, terahertz detectors \cite{liu2009development, liu2010a, volkov2010spectral, jiang2012lens}, electromagnetic pulse (EMP) \cite{stults2008impulse}, and other UWB applications. 

\begin{figure}[!tbp]
	\centerline{\includegraphics[width=\columnwidth]{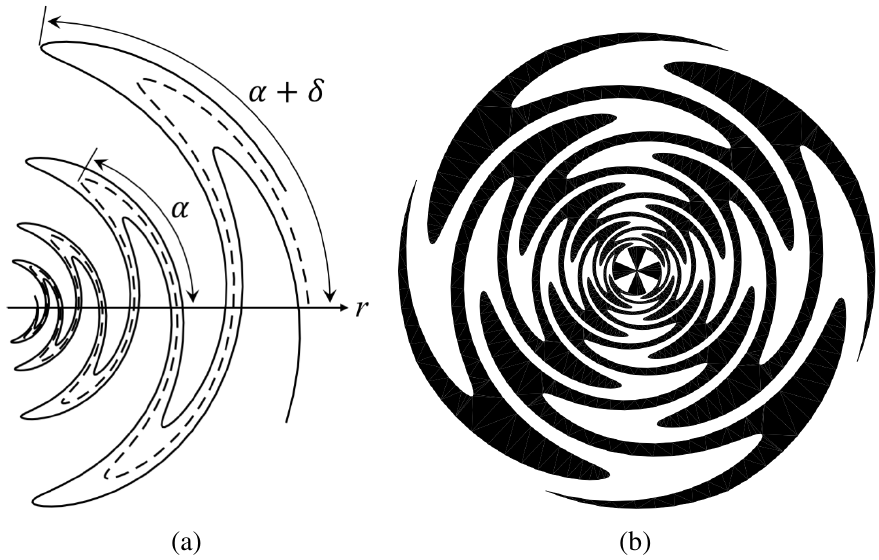}}
	\caption{Designing a sinuous antenna having parameters: $N=\text{4}$ arms (b), $P=\text{8}$ cells, $R_1=\text{5 cm}$, $\tau=\text{0.75}$, $\delta=\text{22.5}^\circ$, and $\alpha=\text{60}^\circ$.}
	\label{fig:1}
\end{figure}

Sinuous antennas are comprised of $N$ arms each made up of $P$ cells where the curve of the $p^\text{th}$ cell is described in polar coordinates ($r$, $\phi$) by
\begin{equation}\label{eq:sin}
\phi = (-1)^{p-1}\alpha_p\sin \left( \frac{\pi \ln(r/R_p)}{\ln(\tau_p)} \right) \pm \delta,
\end{equation}
where $R_{p+1} \le r \le R_p$ \cite{duhamel1987dual}. In (1), $R_p$ controls the outer radius, $\tau_p$ the growth rate i.e., $R_{p+1}=\tau R_p$, and $\alpha_p$ the angular width of the $p^\text{th}$ cell. The curve is then rotated $\pm$ the angle  $\delta$ in order to fill out the arm. In this analysis, four arm ($N=\text{4}$) sinuous antennas are considered with $\alpha$ and $\tau$ kept constant for all cells and $\delta$ set to 22.5$^\circ$ in order to produce self-complementary structures (see Fig. \ref{fig:1}). The self-complementary  condition ensures that the sinuous antenna's input impedance is both real and frequency independent \cite{edwards2012dual}. The antenna is fed by a self-complimentary arrangement of orthogonal bow-tie elements each feeding a set of opposing sinuous arms.

The use of sinuous antennas in polarimetric radar systems \cite{giuli1986polarization} is especially intriguing due to the ability of the four arm sinuous antenna to produce dual polarized radiation over wide bandwidths. Other wideband antenna designs such as quad-ridge horn \cite{blejer1992ultra}, Vivaldi \cite{pochanin2015ultra}, and resistive-vee \cite{sustman2013a} antennas provide similar capabilities. However, they require relatively large and often complex three dimensional structures in order to produce orthogonal senses of polarization. Alternatively, the sinuous antenna may be implemented as a planar structure. The combination of polarimetric capabilities with a low profile make the sinuous antenna attractive to close-in sensing applications such as ground penetrating radar (GPR). 

Although sinuous antennas provide the desirable properties described above, research has demonstrated that the antennas suffer from resonances which degrade performance. The sharp ends produced by the outer truncation of the sinuous antenna have been shown to resonate when their length is approximately $\lambda/2$ which produces both pattern distortion and ringing in the time domain \cite{saini1996sinuous, kang2013experimental, kang2015ends, kang2015modification}. Further, log-periodic resonances occurring internal to the sinuous arms have been observed and shown to produce additional ringing as well as deleterious effects on gain smoothness, polarization purity, and group delay \cite{kang2015modification, kang2016polarization}.

Such resonances may reduce the effectiveness of sinuous antennas, particularly when applied to close-in sensing applications e.g., GPR. However, techniques for their mitigation have been presented in the literature. The sharp ends produced at the outer truncation of the sinuous antenna can be removed in order to prevent them from resonating \cite{saini1996sinuous, kang2013experimental, kang2015ends, fang2015quad, alotaibi2015cavity, kang2015modification, gawande2011towards, kang2016polarization}. Similarly, in \cite{kang2015modification, kang2016polarization} the sinuous cell tips were clipped along the antenna arm in order to mitigate the log-periodic resonances. While these techniques have been successful, they destroy the self-complimentary nature of the antenna which reduces both elegance and frequency independence. Further, as will be discussed later, the technique presented in \cite{kang2015modification, kang2016polarization} may be viewed as an approximation for a sinuous antenna with a smaller angular width ($\alpha$).

Design equations in \cite{duhamel1987dual} relate $\alpha$ to both the beamwidth and bandwidth of the sinuous antenna. The low frequency cutoff of the sinuous antenna is dependent on $\alpha$ by 
\begin{equation}\label{eq:lf}
\lambda_L/4 = R_1(\alpha + \delta),
\end{equation}
where $\alpha$ and $\delta$ are in radians \cite{duhamel1987dual}.
Therefore, when designing a compact antenna for lower frequencies of operation (e.g., GPR), antenna designers may be motivated to choose large values for $\alpha$. However, research presented herein will show that the aforementioned log-periodic resonances are directly correlated to the choice of $\alpha$ with larger values accentuating the resonances. Thus, when selecting a value of $\alpha$, a trade-off between resonance severity and low frequency cutoff must be made. Similar guidance is provided in \cite{duhamel1993frequency} which suggests $\alpha + \delta$ be kept to $\le\text{70}^\circ$ ``to ensure good efficiency and gain performance without dropouts over the frequency band.'' However, to the knowledge of the authors, no additional analysis on the correlation between design parameters and the log-periodic resonances is provided in the literature.

In this paper, section II investigates the correlation between the design parameters $\alpha$ and $\tau$ and the periodic resonances. It will be shown that the periodic resonances may be eliminated by proper choice of design values alone without any additional modification e.g., clipping the cell tips. Section III presents a novel method for truncation of the sinuous antenna which mitigates the resonance due to the sharp ends of the antenna without destroying self-complementariness. Finally, section IV presents the results of an improved sinuous antenna design which implements the methods developed in the previous sections.

\section{Log-periodic Resonances}\label{sec:res}
\begin{figure}[]
	\centerline{\includegraphics[width=\columnwidth]{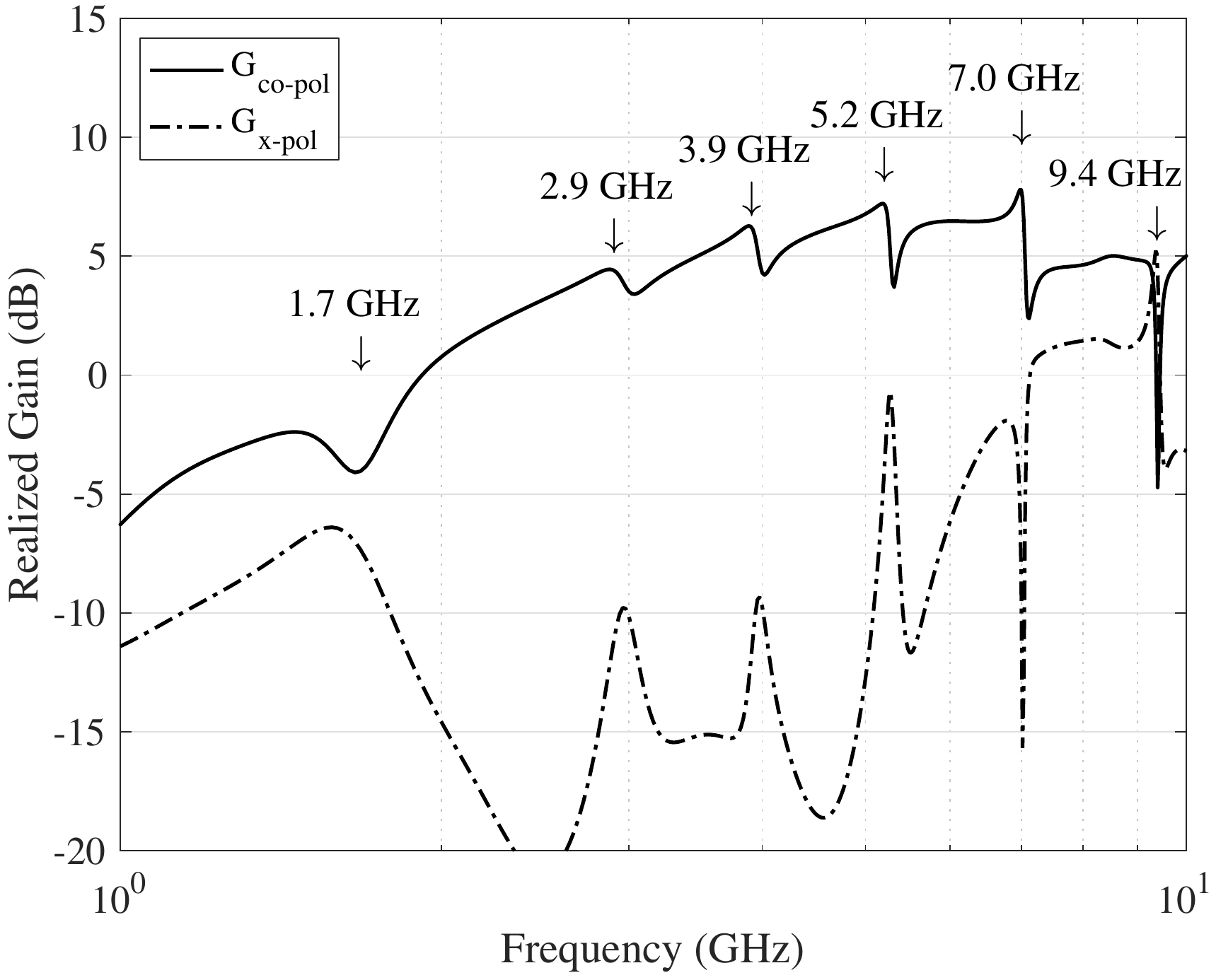}}
	\caption{Realized gain (logarithmic frequency scale) of sinuous antenna having parameters: $N=\text{4}$ arms, $P=\text{8}$ cells, $R_1=\text{5 cm}$, $\tau = \text{0.75}$, $\delta = \text{22.5}^\circ$, and $\alpha = \text{60}^\circ$.}
	\label{fig:p102}
\end{figure}

In order to investigate the log-periodic resonances known to occur in sinuous antenna radiation, full-wave electromagnetic analysis of the antenna described by Fig. \ref{fig:1}b\footnote{
	The sinuous antenna described in Fig. \ref{fig:1}b has parameters similar to the unmodified sinuous antennas analyzed in \cite{kang2015modification} and \cite{kang2016polarization}.
} was conducted using the CST Microwave Studio \cite{cst} frequency domain solver with adaptive meshing\footnote{
    Adaptive meshing was applied at 4 GHz with the resulting mesh utilized for all other frequency samples. Mesh sizes ranged from 211,704 to 366,184 elements depending on the sinuous antenna design parameters (spot checks were made at higher mesh densities to confirm accuracy). Corresponding simulation times varied from 18 to 70 hours for 1000 frequency samples on a 3.7 GHz quad-core Intel Xeon E5-1620 v2 CPU.
}.
Both pairs of opposing sinuous arms were terminated by an ideal port set to the theoretical impedance of 267 $\Omega$ \cite{edwards2012dual}. A single pair was then driven, with the other pair remaining matched, in order to produce linearly polarized radiation. The resulting realized gain is shown in Fig. \ref{fig:p102} and displays prominent resonances. The resonance at approximately 1.7 GHz is attributed to the sharp-ends of the antenna \cite{kang2015modification} and will be further discussed in section \ref{sec:trunc}. The additional resonances starting at 2.9 GHz are the topic of this section. 

Since the values of both $\alpha$ and $\tau$ remain constant for each cell in the sinuous antenna analyzed, the antenna is considered to be a log-periodic structure \cite{duhamel1987dual}. Thus, the radiated fields at frequency $f$ will repeat, since the structure repeats (scaled in size), at frequencies $\tau^nf$ where $n$ is an integer \cite{duhamel1957broadband}. As expected, the resonances observed in Fig. \ref{fig:p102} after 1.7 GHz are logarithmically periodic with period $|\ln\tau|$ indicating the physical cause of a resonance repeats for each cell. This is consistent with the work presented in \cite{kang2015modification} which showed each resonance may be attributed to a specific cell. In this analysis, the correlation between the sinuous antenna design parameters and the log-periodic resonances is investigated.

\subsection{Parametric Study}
A parametric study of both the angular width $\alpha$ and expansion ratio $\tau$ was conducted in order to determine correlation with the log-periodic resonances. For this study, the value of $\alpha$ was swept from 35$^\circ$ to 65$^\circ$ in 5$^\circ$ increments for three values of $\tau$ while maintaining all other sinuous design parameters constant (see summary in Table \ref{tbl:param}). Simulated current distributions, presented in previous work, indicated strong interactions between adjacent sinuous arms at resonance frequencies \cite{kang2015modification}. Therefore, in an attempt to mitigate such interactions, values of $\alpha$ less than 60$^\circ$ were investigated because decreasing $\alpha$ results in less interleaving of adjacent sinuous arms.

\begin{table}
	\caption{Sinuous Antenna Design Parameters for Parametric Sweep}
	\setlength{\tabcolsep}{3pt}
	\begin{tabular}{p{90pt}p{30pt}p{115pt}}
		\toprule
		Parameter & Symbol & Value(s) \\
		\midrule
		Number of Arms & $N$ & 4 \\
		Number of Sinuous Cells & $P$ & 8, 12, 16 \\
		Angular Width & $\alpha$ & 35$^\circ$, 40$^\circ$, 45$^\circ$, 50$^\circ$, 55$^\circ$, 60$^\circ$, 65$^\circ$\\
		Outer Radius & $R_1$ & 5 cm \\
		Inner Radius (feed) & $R_{in}$ & 0.5 cm \\
		Expansion Ratio & $\tau$ & 0.75, 0.825, 0.866 \\
		Arm Sweep Angle & $\delta$ & 22.5$^\circ$ \\
		\bottomrule
	\end{tabular}
	\label{tbl:param}
\end{table}

\begin{figure*}[]
	\centerline{\includegraphics[width=\textwidth]{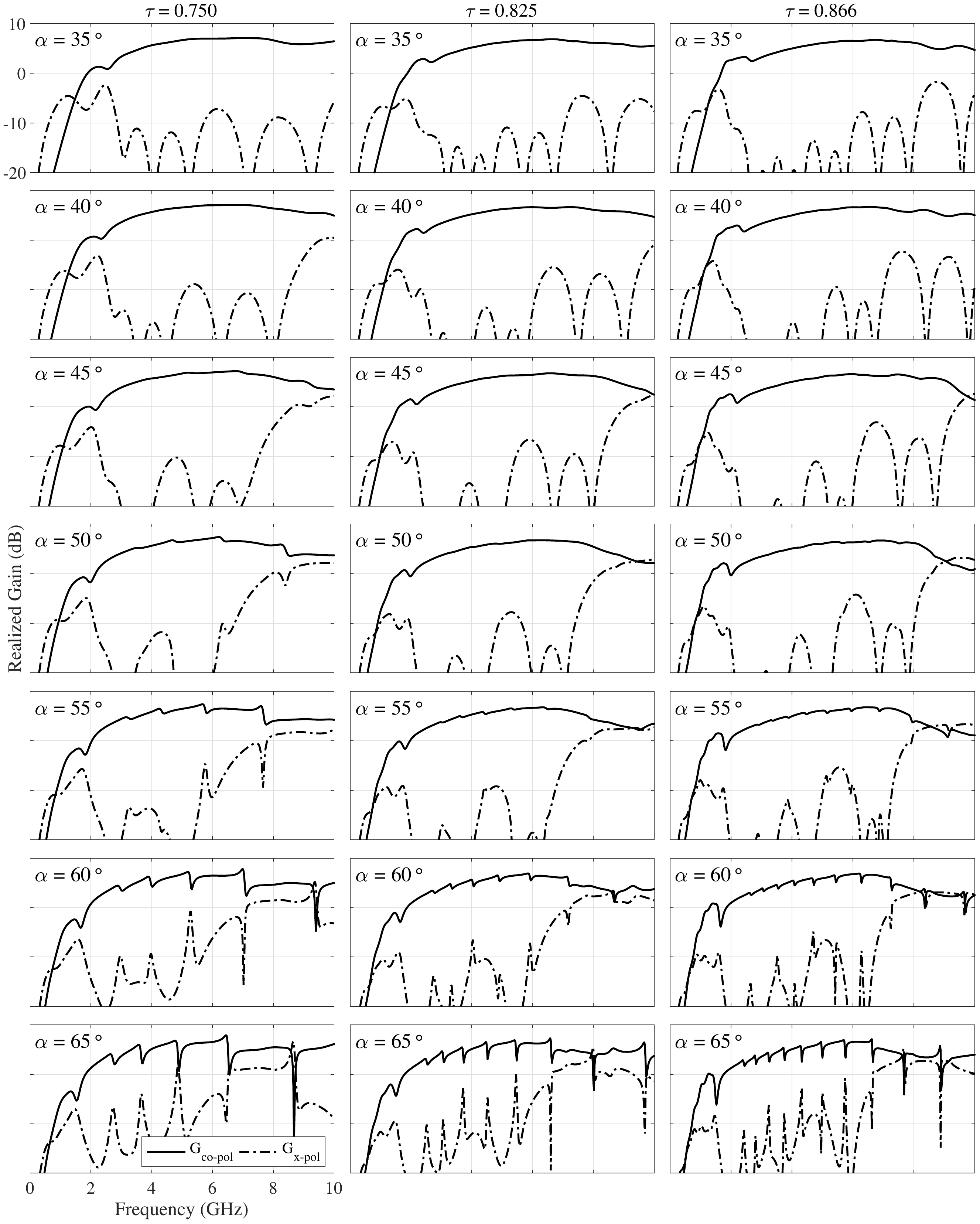}}
	\caption{Realized gains of sinuous antennas with values of $\alpha$ between 35$^\circ$ and 65$^\circ$: (a) $\tau=0.75$ and $P=8$, (b) $\tau=0.825$ and $P=12$, (c) $\tau=0.866$ and $P=16$.}
	\label{fig:p103}
\end{figure*}

\begin{figure}[]
	\centerline{\includegraphics[width=\columnwidth]{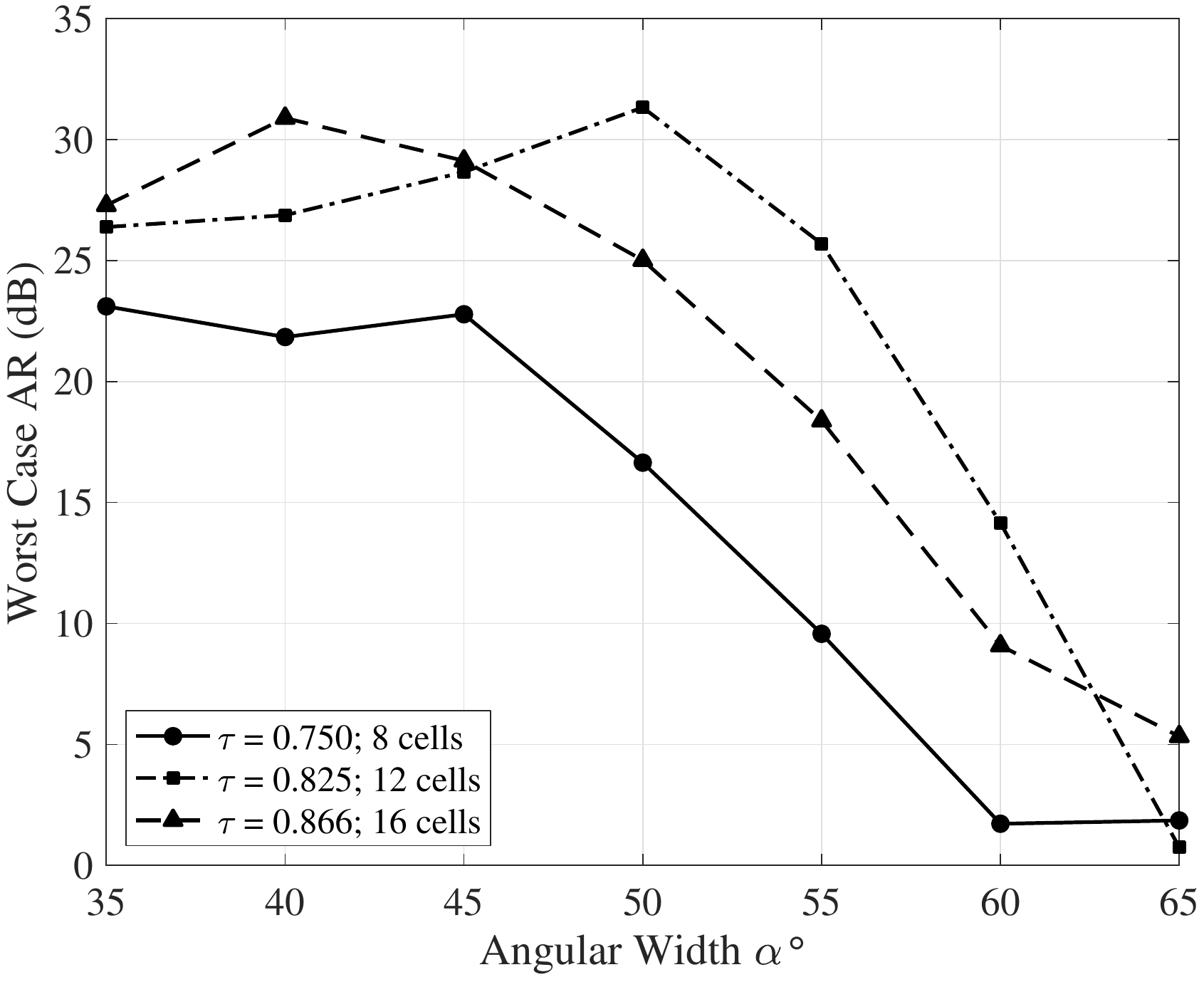}}
	\caption{Worst case axial ratio (AR) due to log-periodic resonances for the parametric study.}
	\label{fig:p105}
\end{figure}

\begin{figure}[]
	\centerline{\includegraphics[width=\columnwidth]{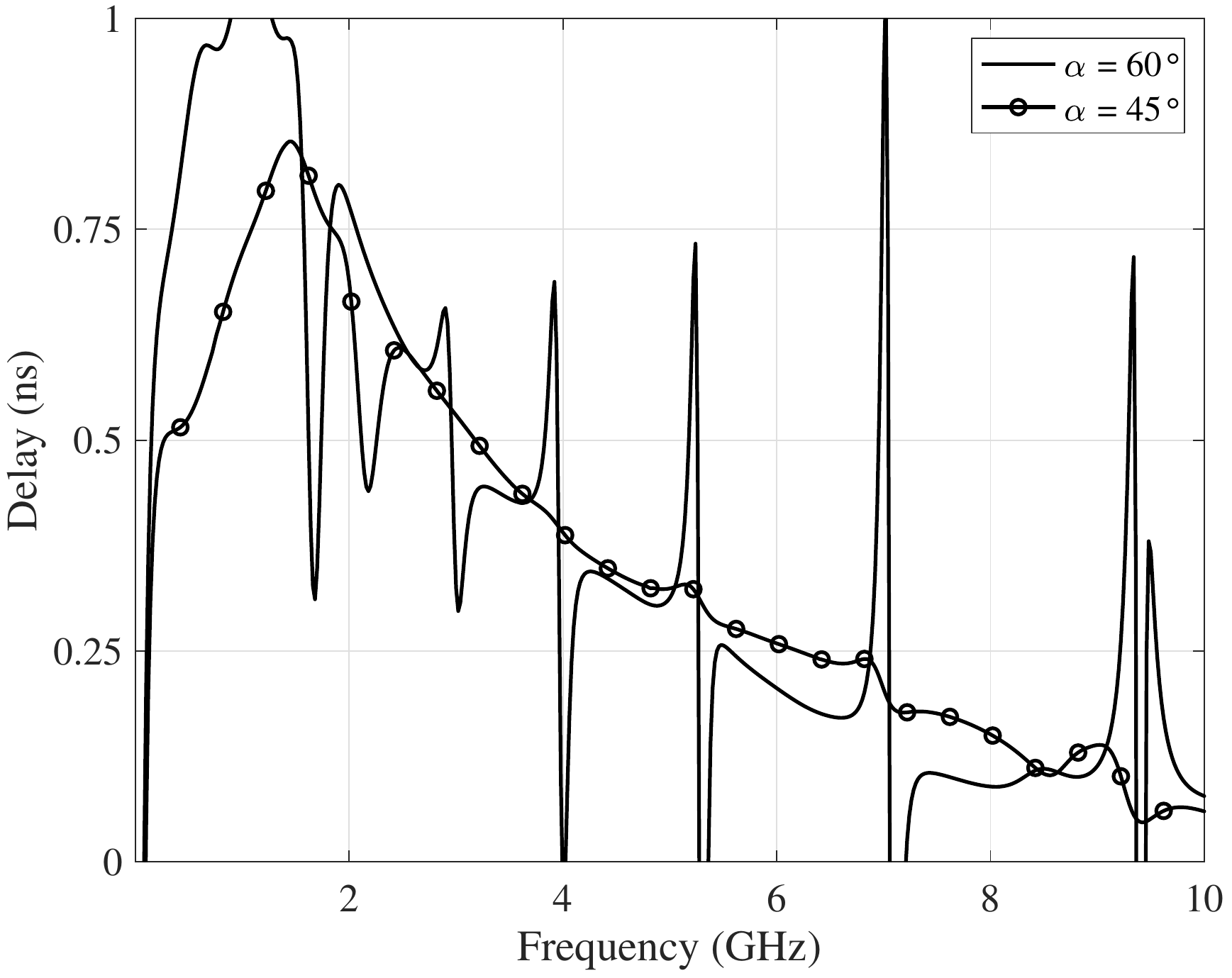}}
	\caption{Comparison of group delay for 8 cell ($\tau=$ 0.75) sinuous antennas having $\alpha$ equal to 45$^\circ$ and 60$^\circ$.}
	\label{fig:p104}
\end{figure}

Simulated boresight realized gains for each antenna with a different pair of $\alpha$ and $\tau$ are shown in Fig. \ref{fig:p103}. From these results, it is evident that decreasing $\alpha$ causes a reduction in the amplitude of the resonances (including that due to the sharp ends). In fact, the log-periodic resonances are no longer visible at approximately $\alpha\le\text{45}^\circ$ for all values of $\tau$. Additionally, improvements to both axial ratio (AR) and group delay, similar to those obtained by cell trimming in \cite{kang2016polarization}, were also observed. Fig. \ref{fig:p105} shows worst case AR computed over the band containing the log-periodic resonances (not including effects of the sharp end resonance) vs. $\alpha$ for each value of $\tau$. As can be seen, mitigation of the log-periodic resonances (smaller $\alpha$) results in a 20 dB improvement in AR (increased linear polarization purity). This is due to the tendency of the resonances to produce elliptically polarized radiation. In addition to the AR, the group delay (shown in Fig. \ref{fig:p104}) exhibits increased smoothness\footnote{
	Note, that the dip in the $\alpha=\text{45}^\circ$ group delay, shown in Fig. \ref{fig:p104}, is due to the resonance caused by the antenna's sharp ends which were not removed for this analysis.
} with decreased $\alpha$ which may be important for dispersion compensation in wideband imaging applications e.g., GPR. 

\begin{figure}[]
	\centerline{\includegraphics[width=\columnwidth]{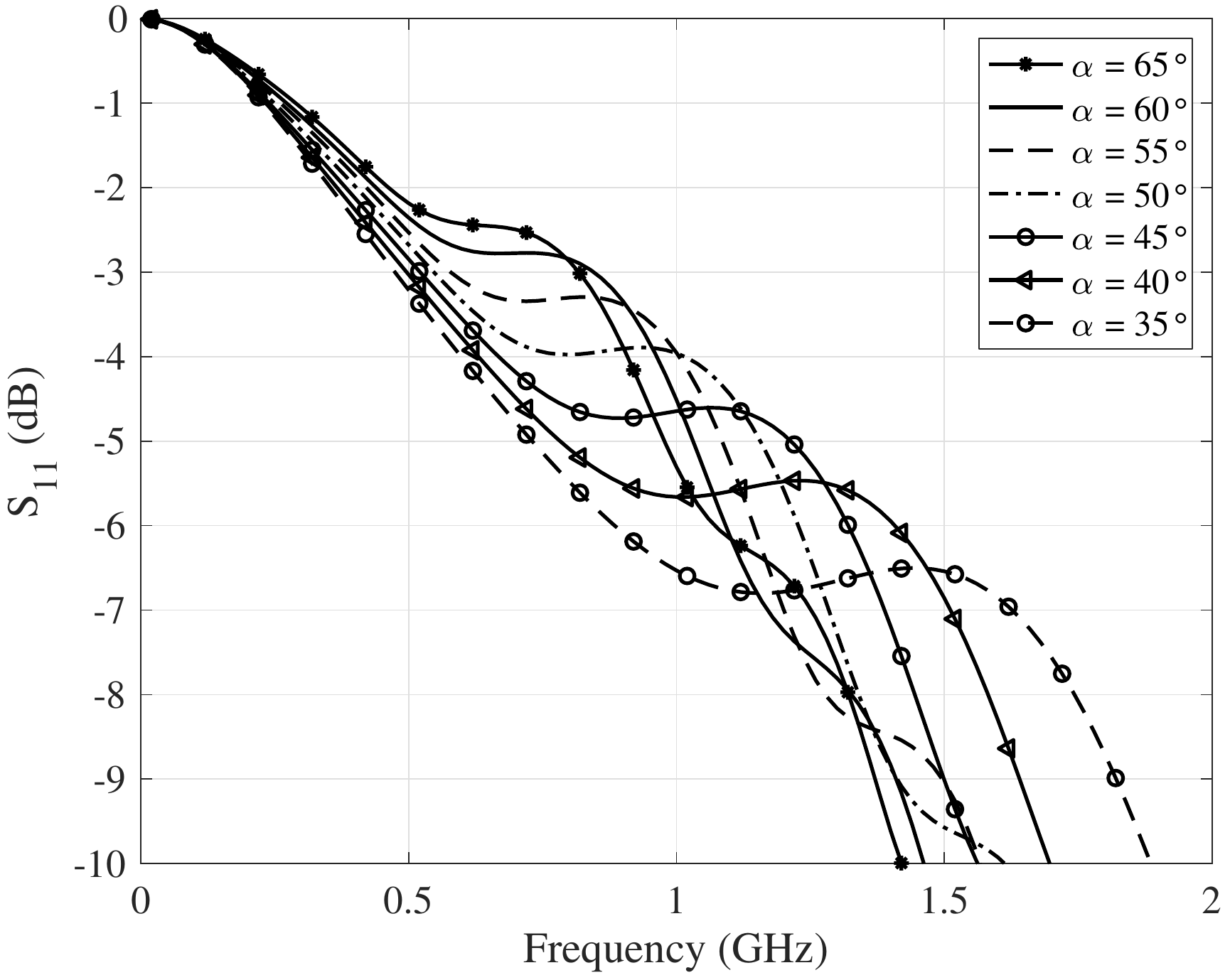}}
	\caption{Comparison of $S_{11}$ for 8 cell ($\tau=$ 0.75) sinuous antennas having $\alpha$ between 40$^\circ$ and 60$^\circ$.}
	\label{fig:p106}
\end{figure}

Although decreasing $\alpha$ to $\le$ 45$^\circ$ results in many benefits to the radiation, the overall length of the sinuous antenna arms also decreased thus negatively impacting the low frequency operation of the antenna. The low frequency $S_{11}$ (matched to an ideal 267 $\Omega$ port) of the simulated 8 cell antennas is shown in Fig. \ref{fig:p106}. As can be seen, reducing $\alpha$ from 65$^\circ$ to 35$^\circ$ resulted 500 MHz less bandwidth with $S_{11} \le -\text{10 dB}$. Therefore, when selecting the design parameter $\alpha$, a trade-off must be made between resonance mitigation and low frequency operation.

\subsection{Gain Smoothness Metric}
The initial results indicated a strong correlation between $\alpha$ and the log-periodic resonances. However, the analysis resulted in a large amount of data which motivated the development of a {\em gain smoothness metric} for determining the success of each parameter combination at mitigating the log-periodic resonances. Since changes in the design parameters resulted in gain variations independent of the resonances, the deviation from a fixed value/line could not be reliably used as a metric. Hence, the metric developed is based on the variation of the realized gain from an $N$-point simple moving average (SMA) computed as
\begin{equation}
G_{SMA}(f_n) = \frac{1}{N}\sum_{k=n-N/2}^{n+N/2}|G(f_k)|,
\end{equation}
where $N=$ 31 (600 MHz wide). More specifically, the metric $M$ is the root mean squared (RMS) error between the co-pol realized gain and its SMA as
\begin{equation}
M =  \sqrt{ \frac{1}{N} \sum_{f_n = f_{start}}^{f_{end}} \big ( |G(f_n)| - G_{SMA}(f_n) \big )^2},
\end{equation}
where in this case $f_{start} \approx \text{3}$ GHz and $f_{stop} = \text{10}$ GHz. The actual starting frequency was adjusted for each design to ensure all the log-periodic resonances were included and not the resonance due to the sharp end. 

\begin{figure}[]
	\centerline{\includegraphics[width=\columnwidth]{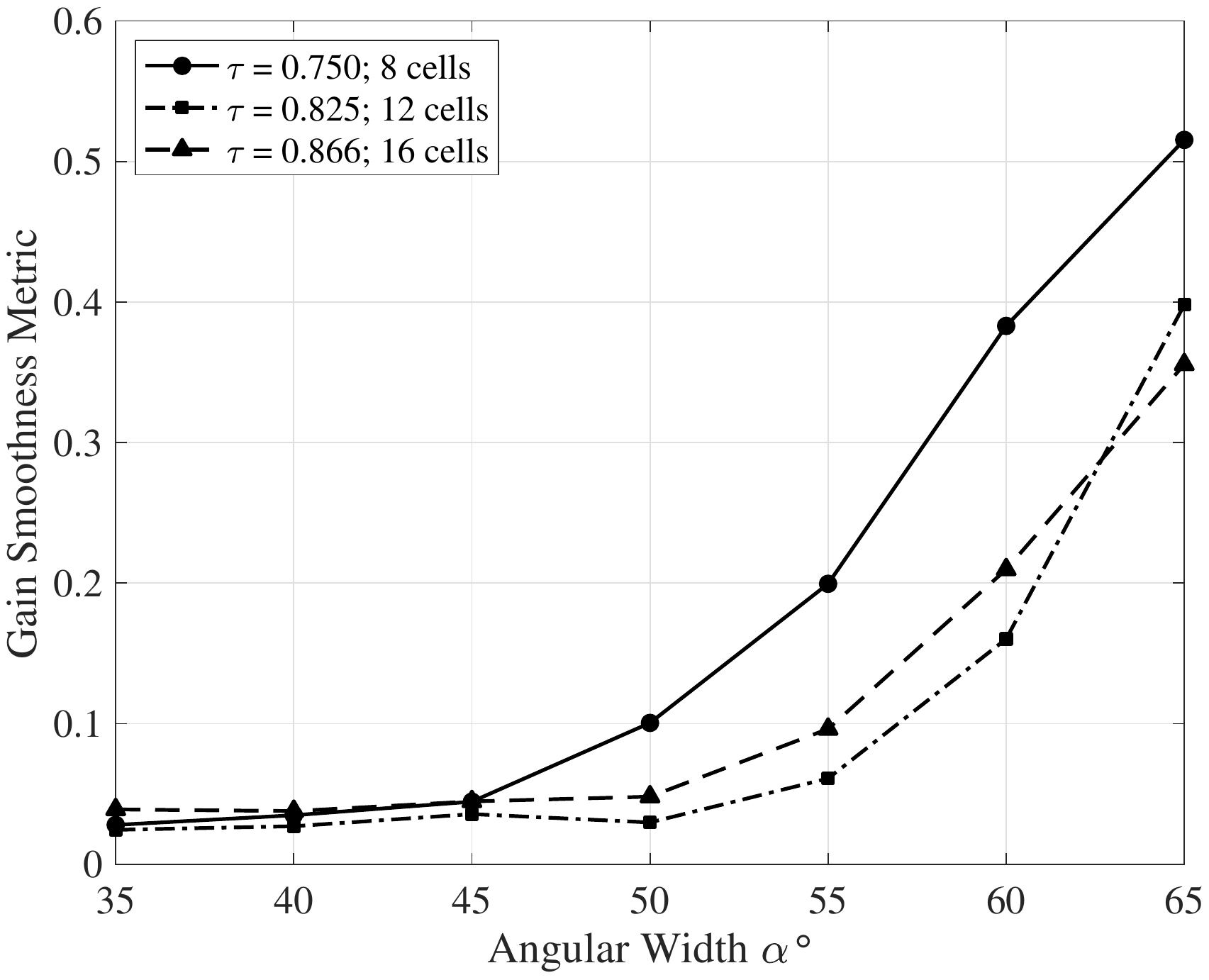}}
	\caption{Gain smoothness metric $M$ computed for the parametric study which shows distortion of gain over frequency increases with $\alpha$.}
	\label{fig:p107}
\end{figure}

The results of the developed metric are shown in Fig. \ref{fig:p107} and indicate that the gain {\em smoothness} converges at approximately $\alpha=\text{45}^\circ$ for all values of $\tau$. Further, the performance of $\alpha$ values larger than 45$^\circ$ is noticeably better for the $\tau= \text{0.825}$ and $\tau=\text{0.866}$ cases compared to the $\tau=\text{0.75}$ case, with $\tau=\text{0.825}$ giving the best results. Therefore, if  $\alpha$ values larger than 45$^\circ$ are desired, e.g., for low frequency applications, the value of $\tau$ may be optimized to improve performance.

While the results clearly indicate the log-periodic resonances in sinuous antennas may be mitigated by proper choice of design parameters, the resonance due to the sharp ends is still present as illustrated by the gains shown in Fig. \ref{fig:p103}. The truncation of the antenna must also be considered in order to obtain resonance free radiation. 

\section{Truncation Resonances}\label{sec:trunc}
\begin{figure}[!tbp]
	\centerline{\includegraphics[width=\columnwidth]{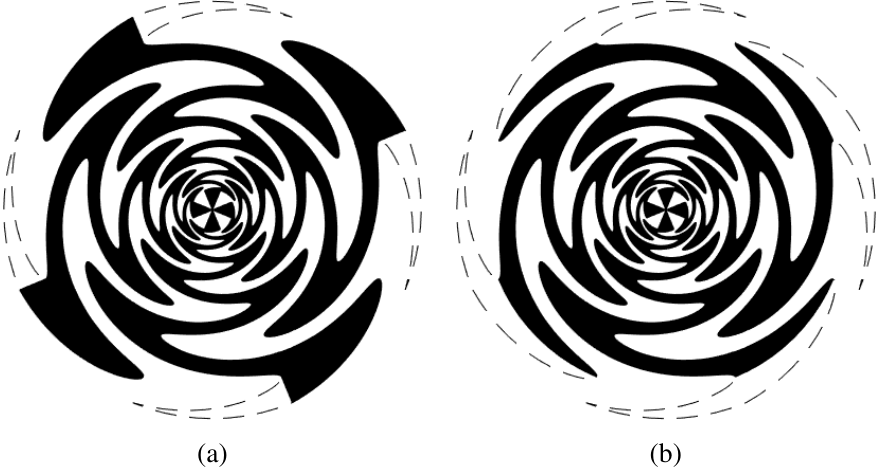}}
	\caption{Sinuous antenna truncation techniques to avoid sharp ends (a) traditional with sharp end clipped, (b) new proposed method with truncation radius moved to the cell tips. Sinuous parameters: $\alpha=\text{45}^\circ$ and $\tau=\text{0.75}$.}
	\label{fig:trunc}
\end{figure}

The outer truncation radius $R_T$ of a sinuous antenna controls the antenna's lowest frequency of operation by the approximate relation
\begin{equation}\label{eq:low}
f_{lo} = \frac{v}{4 R_T (\alpha + \delta)},
\end{equation}
where $v$ is the wave velocity and $\alpha$ and $\delta$ are specified in radians \cite{duhamel1987dual}. Traditionally, the sinuous antenna has been truncated by a circle with radius equal to that of the outermost sinuous cell ($R_T=R_1$) as illustrated by Fig. \ref{fig:1}b. This method of truncation results in a ``sharp end'' at the outer radius of the antenna which produces a resonance that is responsible for the dip in gain observed at 1.7 GHz in Fig. \ref{fig:p102}. While the severity of this resonance is reduced by choosing smaller values of $\alpha$ (see Fig. \ref{fig:p103}), the effects are not removed completely and may be undesirable for some applications. Thus, in order to prevent this resonance, the sharp end is often removed by clipping \cite{saini1996sinuous, kang2013experimental, kang2015ends, fang2015quad, alotaibi2015cavity, kang2015modification, gawande2011towards, kang2016polarization} as shown in Fig. \ref{fig:trunc}a. To the authors' knowledge, no other forms of sinuous antenna truncation have been discussed in the literature.

\subsection{Novel Truncation Method}\label{sec:novel}
An alternative method for the truncation of sinuous antennas that does not produce resonant sharp ends is shown in Fig. \ref{fig:trunc}b. In this method the circular truncation is moved to the tip of the outermost cell thereby preventing the sharp end while maintaining self-complementariness after truncation. This is accomplished by setting the truncation radius $R_T$ to $\sqrt{\tau}R_1$ instead of $R_1$.

\begin{figure}[]
	\centerline{\includegraphics[width=\columnwidth]{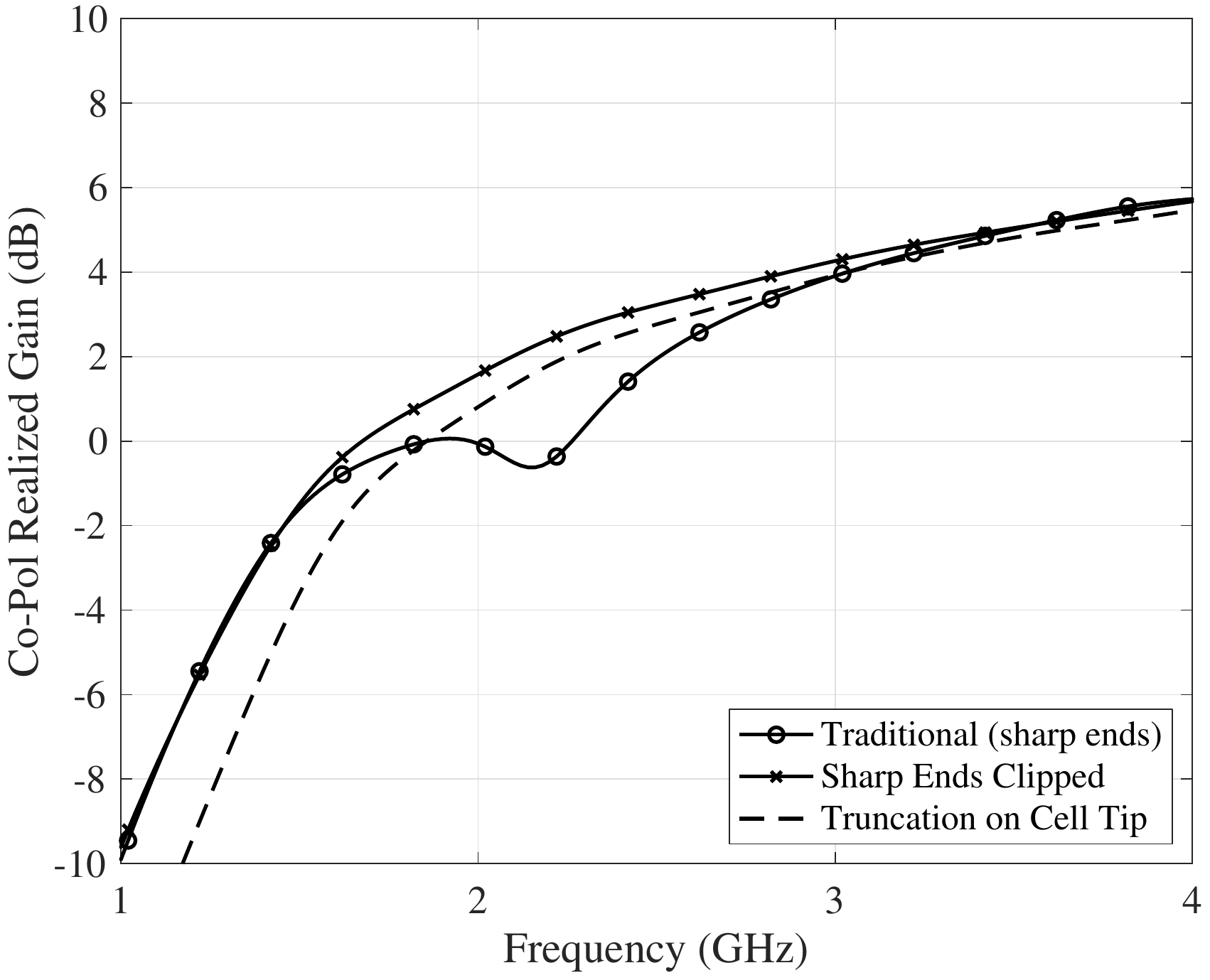}}
	\caption{Comparison of co-polarized realized gain at low frequency for sinuous antennas ($\alpha=45^\circ$ and $\tau=0.75$) with different truncation methods.}
	\label{fig:p111}
\end{figure}

Simulated co-pol realized gains of an $8$ cell $\alpha=\text{45}^\circ$ sinuous antenna truncated by the traditional method with sharp ends,  traditional with sharp ends clipped (Fig. \ref{fig:trunc}a), and the novel tip truncation method (Fig. \ref{fig:trunc}b) are presented in Fig. \ref{fig:p111}. The results show both methods for truncating the antenna without sharp ends successfully mitigate the dip in gain caused by the resonance resulting in an $\sim \text{3}$ dB improvement. However, the low frequency realized gain of the novel tip truncated sinuous antenna is lower. This is due to the reduction in antenna size by moving the truncation inward on the antenna which results in less usable bandwidth at low frequency. In order to maintain the same outer radius, and thus similar low frequency performance, as a traditionally truncated sinuous antenna, $\tau$ and $R_1$ must be appropriately scaled. 

\section{Improved Sinuous Antenna}
\begin{figure}[!tbp]
	\centerline{\includegraphics[width=\columnwidth]{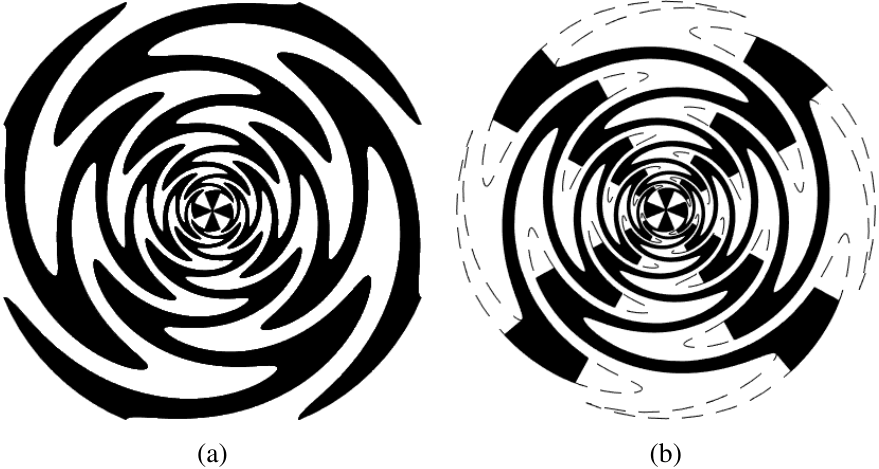}}
	\caption{Sinuous antenna designs modified in order to mitigate the resonances (a) improved design, (b) ends trimming technique.}
	\label{fig:10}
\end{figure}

Building on the results presented in Sections \ref{sec:res} and \ref{sec:trunc}, a sinuous antenna was designed in an effort to produce similar performance as the reference antenna illustrated in Fig. \ref{fig:1}b while mitigating both log-periodic and truncation related resonances. The value of $\alpha$ is chosen as 45$^\circ$ in order to prevent log-periodic resonances while the outer region is truncated by the method proposed in Section \ref{sec:novel}. The antenna in Fig. \ref{fig:trunc}a, $\alpha=\text{45}^\circ$ and sharp ends clipped, could have also been used to provide similar resonance mitigation; however, a fully self-complimentary design was desired. The new antenna is shown in Fig. \ref{fig:10}a and the design parameters are listed in Table \ref{tbl:improv}. Both $\tau$ and $R_1$ have been adjusted in order for the truncation radius to be equal to the outer radius of the reference design ($R_T = R_1^{ref} = \text{5 cm}$). In addition, the radius of the bowtie element feed is  kept to a constant 0.5 cm. The number of cells $P$ is increased by one in order to have the same number of full cells, once the circular truncation is applied to the outermost cell tip, as the reference antenna. From the results presented in Fig. \ref{fig:p107}, one might have based the improved design on the 12 cells antenna with 50$^\circ$ angular width; however, for comparison purposes, particularly in the time-domain, the improved design was kept as close to the reference antenna as possible.

\begin{table}
	\caption{Improved Sinuous Antenna Design Parameters}
	\setlength{\tabcolsep}{3pt}
	\centering
	\begin{tabular}{p{90pt}p{30pt}p{50pt}}
		\toprule
		Parameter & Symbol & Value(s) \\
		\midrule
		Number of Arms & $N$ & 4 \\
		Number of Sinuous Cells & $P$ & 9 \\
		Angular Width & $\alpha$ & 45$^\circ$\\
		Outer Radius & $R_1$ & 5.72 cm \\
		Outer Truncation Radius & $R_T$ & $5$ cm \\
		Inner Radius (feed) & $R_{in}$ & 0.5 cm \\
		Expansion Ratio & $\tau$ & 0.7628 \\
		Arm Sweep Angle & $\delta$ & 22.5$^\circ$ \\
		\bottomrule
	\end{tabular}
	\label{tbl:improv}
\end{table}

For further comparison, the reference antenna was trimmed according the methods proposed in \cite{kang2013experimental} (shown in Fig. \ref{fig:10}b) in order to mitigate the resonances. This required the removal of the sharp ends via clipping as well as a 20$^\circ$ trim of all the sinuous cell tips. This antenna is compared against the reference as well as the improved design.

\subsection{Frequency Domain Analysis}
\begin{figure}[]
	\centerline{\includegraphics[width=\columnwidth]{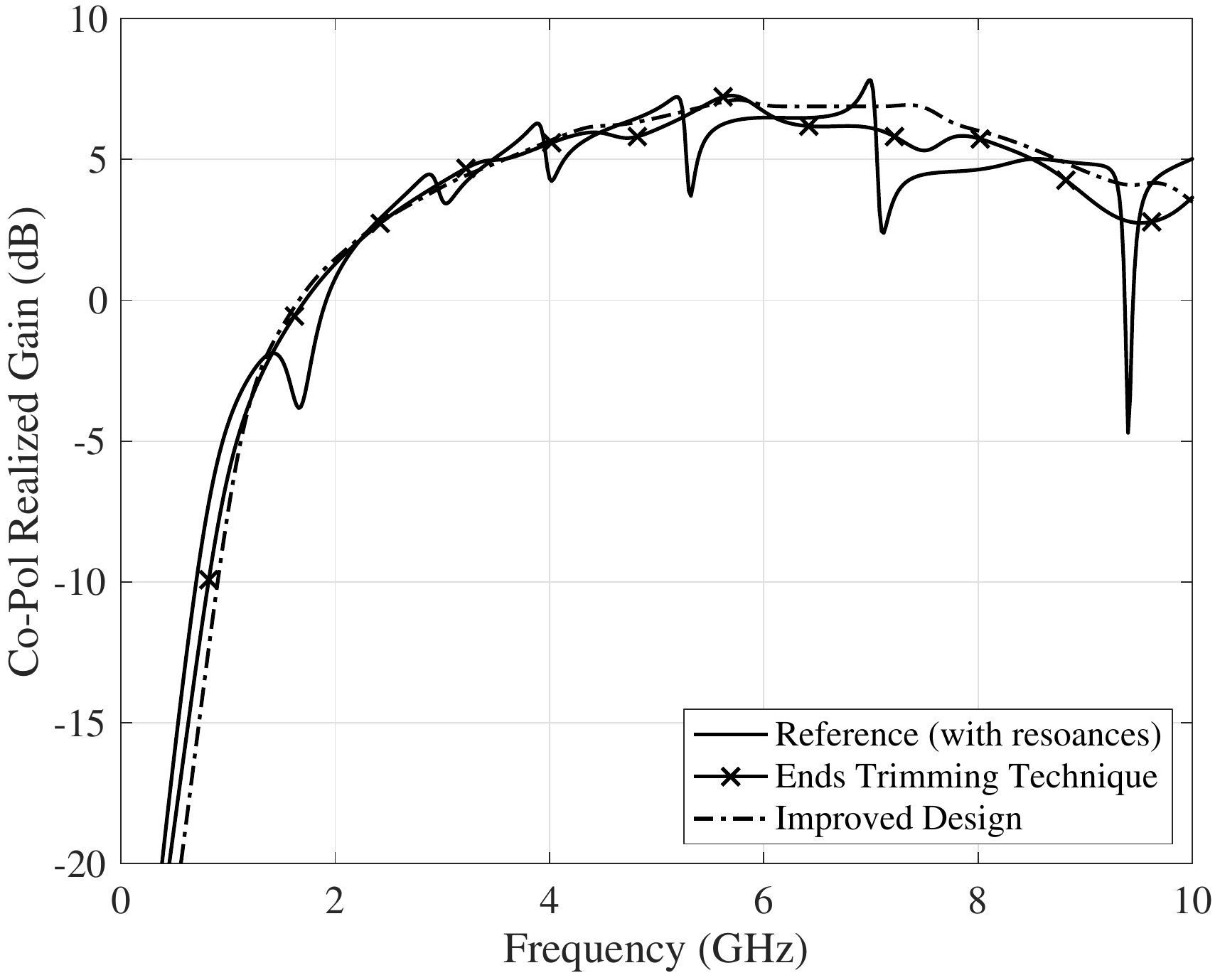}}
	\caption{Comparison of co-polarized realized gain for traditional sinuous with resonances and modified versions that mitigate the resonances.}
	\label{fig:p114}
\end{figure}

\begin{figure}[]
	\centerline{\includegraphics[width=\columnwidth]{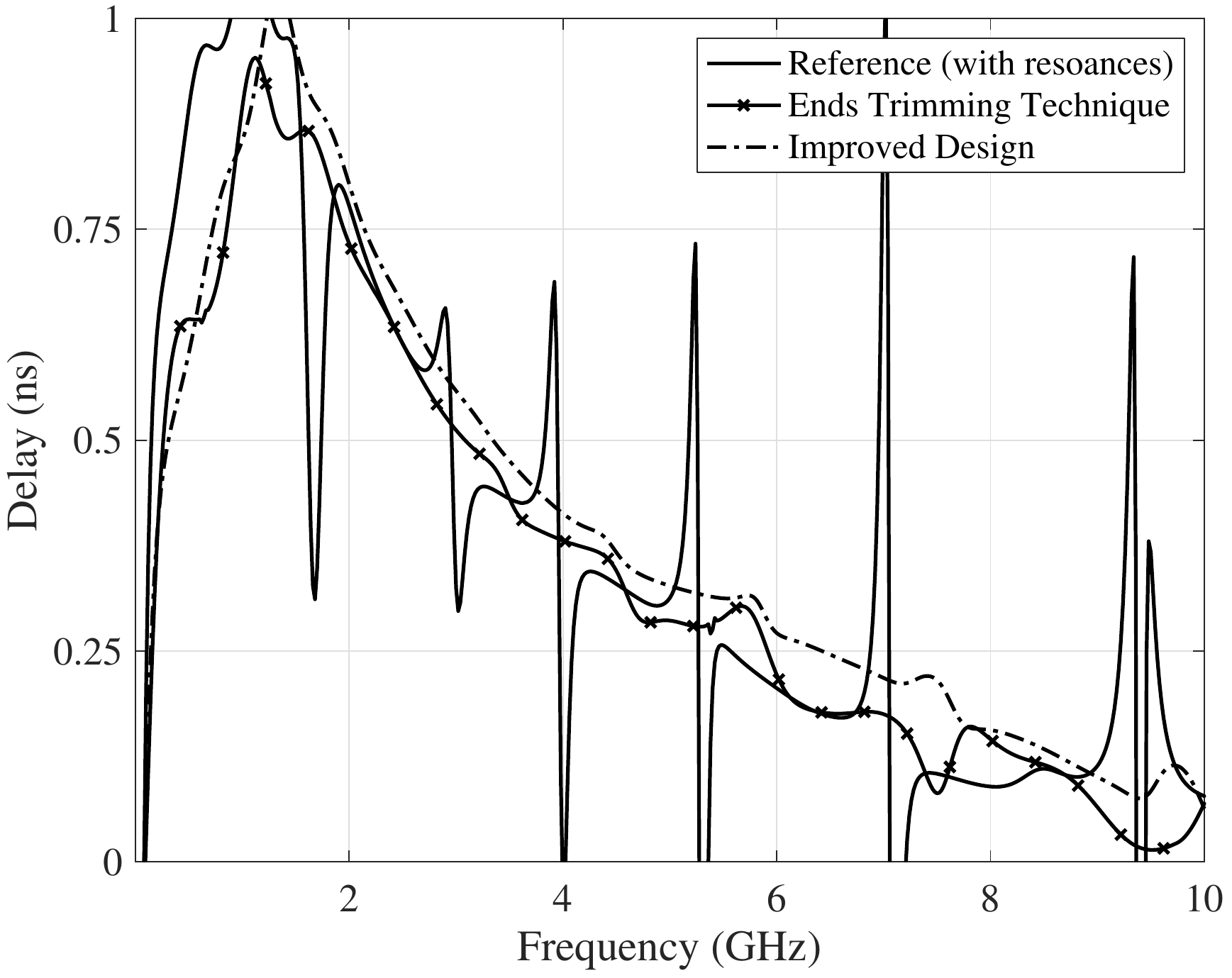}}
	\caption{Comparison of group delay for traditional sinuous with resonances and modified versions that mitigate the resonances.}
	\label{fig:p117}
\end{figure}

\begin{figure}[]
	\centerline{\includegraphics[width=\columnwidth]{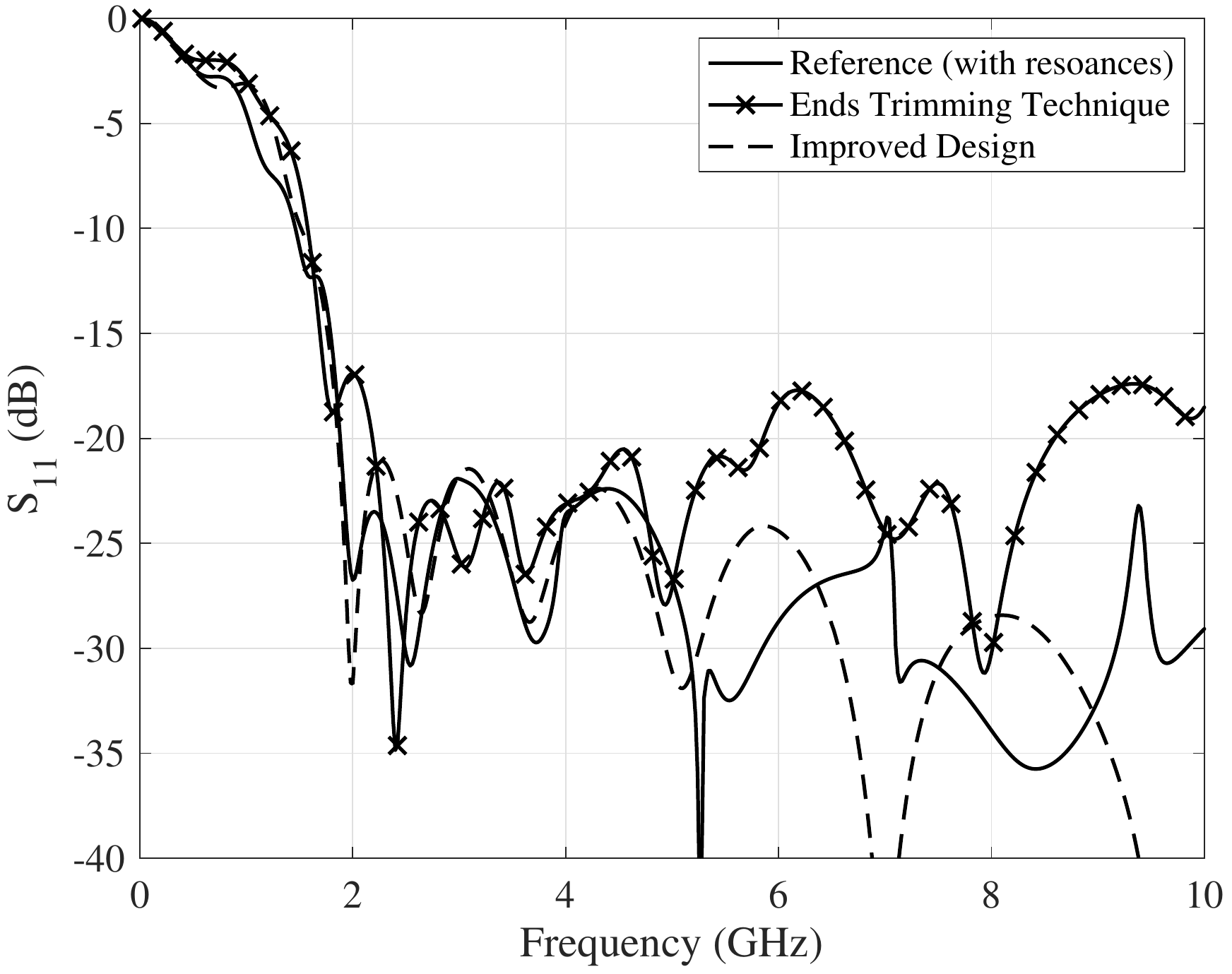}}
	\caption{Comparison of match ($S_{11}$) for traditional sinuous with resonances and modified versions that mitigate the resonances.}
	\label{fig:p115}
\end{figure}

The simulated boresight co-polarized realized gains of the three antennas are shown in Fig. \ref{fig:p114}. The results show that the improved design successfully mitigates both the log-periodic and truncation resonances while providing gain of similar (sometimes higher) magnitude. In comparison, the trimmed design is also able to mitigate the resonances but the realized gain is not as smooth. Similarly, the group delay is significantly smoother with the removal of the resonances as shown in Fig. \ref{fig:p117}. The simulated match ($S_{11}$) of the antennas to an ideal 267 $\Omega$ port is shown in Fig. \ref{fig:p115}. The different designs all have $S_{11} \le -\text{10 dB}$ starting at approximately 1.5 GHz ($\pm\,\text{50 MHz}$) with the improved design having an overall better match than the trimmed version which is no longer self-complimentary due to trimming. 

\subsection{Time Domain Analysis}
\begin{figure}[]
	\centerline{\includegraphics[width=\columnwidth]{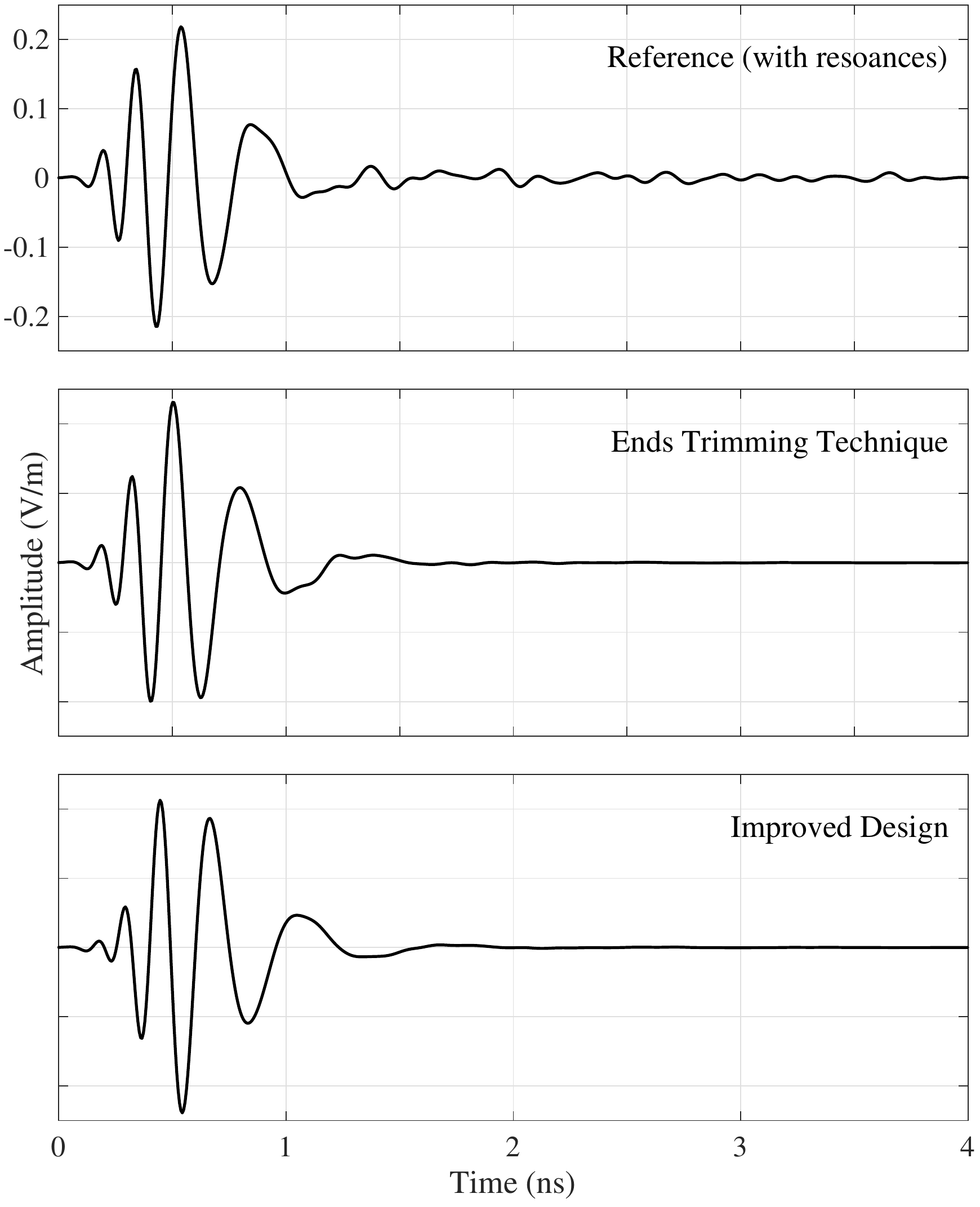}}
	\caption{Comparison of radiated pulses (time shifted to zero) for traditional sinuous with resonances and modified versions that mitigate the resonances.}
	\label{fig:p108}
\end{figure}

Resonances present in the gain of the sinuous antenna ultimately result in ringing when the antenna is used in pulsed type applications which can be particularly troublesome for close in sensing applications such as GPR. Examination of the radiated fields in the time domain is necessary to determine the extent of such ringing. Fig. \ref{fig:p108} shows the radiated pulses (time-shifted to start at $t=\text{0}$) from the three sinuous antennas when driven by an UWB pulse. The excitation pulse used was a Differentiated Gaussian which is defined by
\begin{equation}\label{eq:pulse}
v_{pulse}(t) = -v_{peak}\frac{(t - \mu)}{\sigma} \exp \left [ 0.5 - \frac{(t - \mu)^2}{2 \sigma^2} \right ], 
\end{equation}
where $\mu$ represents an arbitrary time shift and $\sigma$, defined as $2.3548/(2\pi f_{BW})$, controls the width of the pulse. In the presented analysis  $v_{peak}$ was set to 1 V and $f_{BW}$ to 7.5 GHz resulting in peak spectral energy at 3.2 GHz.

As shown in Fig. \ref{fig:p108}, the resonances produce high frequency ringing following the pulse. Mitigation of the resonances significantly reduces this ringing. The improved design has a slightly different pulse shape compared to the other designs which is due to the change in $\tau$. Both the improved and trimmed designs provide similar performance improvements i.e., less ringing. However, the improved design is able to achieve this while maintaining a self-complimentary structure.

\section{Conclusion}
In this paper, an analysis of the relationship between the sinuous antenna design parameters and the log-periodic resonances in the radiation was presented. It was determined that these log-periodic resonances may be eliminated by proper selection of the arm angular width $\alpha$. By choosing a value of $\alpha$ that reduces the interleaving of the arms, the log-periodic resonances were subsequently reduced. The optimal value of $\alpha$ was slightly impacted by the choice of expansion ratio $\tau$ but converged at approximately 45$^\circ$ for all the designs investigated. These results are corroborated by design guidance provided in \cite{duhamel1993frequency}. Mitigation of the log-periodic resonances by proper choice of design parameters provides advantages over other techniques proposed in the literature since the antenna remains self-complimentary and does not require any special modification of the arms. Further, this relationship explains why some researchers have encountered the log-periodic resonances and others have not.

In addition to the log-periodic resonances, a resonance occurs at low frequency due to the sharp end left by the sinuous antenna outer truncation. Such a resonance may be problematic for some applications. It was shown that by moving the truncation radius to the tip of the outermost sinuous cell both the sharp end and the corresponding resonance are eliminated. This novel truncation method has the benefit of maintaining the self-complimentary structure of the antenna as opposed to manual removal of the sharp ends via clipping. 

Finally, a sinuous antenna was designed using these principles and was shown via simulation to provide the desired performance in both the frequency and time domain. More specifically, the antenna produced smooth radiation over a wide band while maintaining a good match to the theoretical input impedance.

\section*{Acknowledgment}
The authors would like to thank Bernd H. Strassner II and John J. Borchardt of Sandia National Laboratories for helpful discussion and feedback during the development of this paper.

\bibliographystyle{IEEEtran}
\bibliography{crocker_scott_sinuous_tap}

\end{document}